# Detection of Cooperative Black Hole Attack in Wireless Ad Hoc Networks

Jaydip Sen
Innovation Lab
Tata Consultancy Services Ltd.
Bengal Intelligent Park, Salt Lake Electronic Complex, Kolkata, INDIA
e-mail: Jaydip.Sen@tcs.com

*Abstract*— A mobile ad hoc network (MANET) is a collection of autonomous nodes that communicate with each other by forming a multi-hop radio network and maintaining connections in a decentralized manner. Security remains a major challenge for these networks due to their features of open medium, dynamically changing topologies, reliance on cooperative algorithms, absence of centralized monitoring points, and lack of clear lines of defense. Protecting the network layer of a MANET from malicious attacks is an important and challenging security issue, since most of the routing protocols for MANETs are vulnerable to various types of attacks. Ad hoc on-demand distance vector routing (AODV) is a very popular routing algorithm. However, it is vulnerable to the well-known black hole attack, where a malicious node falsely advertises good paths to a destination node during the route discovery process but drops all packets in the data forwarding phase. This attack becomes more severe when a group of malicious nodes cooperate each other. The proposed mechanism does not apply any cryptographic primitives on the routing messages. Instead, it protects the network by detecting and reacting to malicious activities of the nodes. Simulation results show that the scheme has a significantly high detection rate with moderate network traffic overhead and computation overhead in the nodes.

## I. INTRODUCTION

A MANET is a collection of wireless hosts that can be rapidly deployed as a multi-hop packet radio network without the aid of any established infrastructure or centralized administrator. Such networks can be used to enable next generation battlefield applications, including situation awareness systems for maneuvering war fighters, and remotely deployed unmanned micro-sensor networks. MANETs have some special characteristic features such as unreliable wireless media (links) used for communication between hosts, constantly changing network topologies and memberships, limited bandwidth, battery, lifetime, and computation power of nodes etc. While these characteristics are essential for the flexibility of MANETs, they introduce specific security concerns that are absent or less severe in wired networks. MANETs are vulnerable to various types of attacks. These include passive eavesdropping, active interfering, impersonation, and denial-of-service. Intrusion prevention measures such as strong authentication and redundant transmission can be used to improve the security of an ad hoc network. However, these techniques can address only a subset of the threats. Moreover, they are costly to implement. The dynamic nature of ad hoc networks requires that prevention techniques should be complemented by detection techniques, which monitor security status of the network and identify malicious behavior.

One of the most critical problems in MANETs is the security vulnerabilities of the routing protocols. A set of nodes in a MANET may be compromised in such a way that it may not be possible to detect their malicious behavior easily. Such nodes can generate new routing messages to advertise non-existent links, provide incorrect link state information, and flood other nodes with routing traffic, thus inflicting Byzantine failure in the network. One of the most widely used routing protocols in MANETs is the *ad hoc on-demand distance vector* (AODV) routing protocol [1]. It is a source initiated on-demand routing protocol. However, AODV is vulnerable to the well-known black hole attack. In [2], the authors have assumed that the black hole nodes in a MANET do not work as a group and have proposed a solution to identify a single black hole. However, their proposed method cannot be applied to identify a cooperative black hole attack involving multiple malicious nodes. In this paper, a mechanism is proposed to identify multiple black hole nodes cooperating as a group in an ad hoc network. The is an extension of our previous work [3], which is a modified AODV protocol that makes use of a *data routing information table* (Section III). Extensive simulations have been done for evaluating the performance of the protocol.

The rest of the paper is organized as follows. Section II discusses some related work in security mechanism in routing for MANETs. Section III gives an overview of AODV protocol and the cooperative black hole attack. Section IV describes the proposed security protocol and the associated algorithm. Section V presents the important results obtained in simulation. Section VI concludes the paper while highlighting some future scope of work.

## II. RELATED WORK

The problem of security and cooperation enforcement has received considerable attention by researchers in the ad hoc network community. In this section, some of these contributions are presented.

The problem of securing the routing layer using cryptographically secure messages is addressed by Hu et al. [4], Papadimitratos and Haas [5], and Sanzgiri et al. [6]. Schemes to handle authentication in ad hoc networks assuming trusted certificate authorities have been proposed by Kong et al. [7]. Hubaux et al. [8] have employed a self-organized PGP-based scheme to authenticate nodes using chains of certificates and transitivity of trust. Stajano et al. [9] have proposed authentication of users by imprinting in analogy to ducklings acknowledging the first moving subject they see as their mother.

In contrast to securing the routing layer of ad hoc networks, some researchers have also focused on simply detecting and reporting misleading routing misbehavior. *Watchdog* and *Pathrater* [10] use observation-based techniques to detect misbehaving nodes, and report observed misbehavior back to the source of the traffic. Pathrater manages trust and route selection based on these reports. This allows nodes to choose better paths along which to route their traffic by routing around the misbehaving nodes. However, the scheme does not punish malicious nodes; instead, they are relieved of their packet forwarding burden.

CONFIDANT [11] detects misbehaving nodes by means of observation and more aggressively informs other nodes of this misbehavior through reports sent around the network. Each node in the network hosts a *monitor* for observations, *reputation records* for first-hand and trusted second-hand reports, *trust records* to control the trust assigned to the received warnings, and a *path manager* used by nodes to adapt their behavior according to reputation information. Subsequent research has found that reputation schemes can be beneficial for fast misbehavior detection, but only when one can deal with false accusations [12].

Researchers have also investigated means of discouraging selfish routing behavior in ad hoc networks, generally through payment schemes [13]. These approaches either require the use of tamper-proof hardware or central bankers to do the accounting securely, both of which may not be appropriate in some truly ad hoc network scenarios. In the per-hop payment scheme proposed by Buttyan et al. [14], the payment units are called *nuglets* and reside in a secure tamper-proof module in each node. The scheme relies on first-hand observations by the nodes. Directly observed positive behavior increases the rating of a node, while directly observed negative behavior decreases it by an amount larger than that is used for positive increments. If the rating of a node dips below the faulty threshold, the node is added to the faulty list. The faulty list is appended to the route request by each node broadcasting it to be used as a list of nodes to be avoided. A route is rated good or bad depending on whether the next hop is on the faulty list. If the next hop of a route is in the faulty list, the route is rated as bad. As a response to misbehavior of a node, all traffic from that node is rejected. A second chance mechanism for redemption employs a timeout after an idle period. After a timeout, the node is removed from the faulty list with its rating remaining unchanged. They have observed that given such a module, increased cooperation is beneficial not only for the entire network but also for individual nodes. The scheme can result in unfairness to some hosts, but its simplicity and performance may be appropriate in some cases.

Bansal et al. [15] have proposed a scheme that relies on first-hand observations. Directly observed positive behavior increases the rating of a node, while directly observed negative behavior decreases it by an amount larger than that is used for positive increments. If the rating of a node dips below the faulty threshold, the node is added to a faulty list. The faulty list is appended to the route request by each node broadcasting it to be used as a list of nodes to be avoided. A route is rated good or bad depending on whether the next hop is on the faulty list. If the next hop of a route is in the faulty list, the route is rated as bad. As a response to misbehavior of a node, all traffic from that node is rejected. A second chance mechanism for redemption employs a timeout after an idle period. After a timeout, the node is removed from the faulty list with its rating remaining unchanged.

In [16], a security architecture for MANETs has been proposed that involves mobile agents. In this scheme, multiple sensors deployed throughout the network collect and merge audit data implementing a cooperative detection algorithm. Sensors are deployed on some of the hosts in the network that monitor the network traffic. The selection of these nodes is based on the connectivity index and a distributed voting algorithm. The detection decisions are taken by mobile agents that migrate their execution and state information between different sensor hosts of the network, and finally return to the originator host with the results. The authors have proposed two different methods of decision-making: collaborative and independent. They argue that independent decision-making by mobile agents is susceptible to single point-of-failure problem and therefore, the collaborative method should be used. The main advantage of this scheme is the restriction of computation-intensive operations of the system to few dynamically elected nodes. However, most of the available mobile agent frameworks are heavyweight and can often be the targets of attacks themselves [17].

Sen et al. have presented a scheme for detection of malicious packet dropping nodes in a MANET [18]. The mechanism is based on local misbehavior detection and flooding of the detection information in a controlled manner in the network so that the malicious node is detected even if moves out a local neighborhood. Sen et al. have also proposed a trust and reputation-based mechanism for misbehavior detection of nodes in a MANET, in which every node computes the reputation information of its neighbors and exchanges the reputation information to make a cooperative detection [19].

Deng, Li and Agarwal [2] have suggested a mechanism of defense against black hole attack in ad hoc networks. In their proposed scheme, as soon as the *RouteReply* packet is received from one of the intermediate nodes, another *RouteRequest* is sent from the source node to a neighbor node of the intermediate node in the path. This is to ensure that such a path exists from the intermediate node to the destination node. For example, let the source node *S* send *RouteRequest* packets and receive *RouteReply* through the intermediate malicious node *M*. The *RouteReply* packet of *M* contains information regarding its next-hop neighbor node. Let it contain information about the neighbor *E*. Then, the source node *S* sends *FurtherRouteRequest* packets to this neighbor node *E*. Node *E* responds by sending a *FurtherRouteReply* packet to source node *S*. Since node *M* is a malicious node, and thus not present in the routing list of node *E*, the *FurtherRouteReply* packet sent by node *E* will not contain a route to the malicious node *M*. But if it contains a route to the destination node *D*, then the new route to the destination through node *E* is selected, and the earlier selected route through node *M* is rejected. While this scheme completely eliminates the black hole attack by a single attacker, it fails completely in identifying a cooperative black hole attack involving multiple malicious nodes.

### III. AODV AND ITS SECURITY PROBLEMS

In this section, a brief overview of the AODV routing protocol is presented and the security threat that are associated with this routing protocol are briefly discussed. More specifically, the cooperative black hole attack on AODV is also described.

#### A. AODV Overview

AODV is a reactive routing protocol that does not require maintenance of routes to destination nodes that are not in active communication. Instead, it allows mobile nodes to quickly obtain routes to new destination nodes. Every mobile node maintains a routing table that stores the next hop node information for a route to the destination node. When a source node wishes to route a packet to a destination node, it uses the specified route if a fresh enough route to the destination node is available in its routing table. If such a route is not available in its cache, the node initiates a route discovery process by broadcasting a *RouteRequest* (RREQ) message to its neighbors. On receiving a RREQ message, the intermediate nodes update their routing tables for a reverse route to the source node. All the receiving nodes that do not have a route to the destination node broadcast the RREQ packet to their neighbors. Intermediate nodes increment the hop count before forwarding the RREQ. A *RouteReply* (RREP) message is sent back to the source node when the RREQ query reaches either the destination node itself or any other intermediate node that has a current route to the destination. As the RREP propagates to the source node, the forward route to the destination is updated by the intermediate nodes receiving a RREP. The RREP message is a unicast message to the source node.

AODV uses sequence numbers to determine the freshness of routing information and to guarantee loop-free routes. In case of multiple routes, a node selects the route with the highest sequence number. If multiple routes have the same sequence number, then the node chooses the route with the shortest hop count. Timers are used to keep the route entries fresh. When a link break occurs, *RouteError* (RERR) packets are propagated along the reverse path to the source invalidating all broken entries in the routing table of the intermediate nodes. AODV also uses periodic *hello* messages to maintain the connectivity of neighboring nodes.

AODV does not incorporate any specific security mechanism, such as strong authentication. Therefore, there is no straightforward mechanism to prevent mischievous behavior of a node such as MAC spoofing, IP spoofing, dropping packets, or altering the contents of the control packets. Protocols like SAR [20] have been developed to secure AODV against certain types of attacks. However, these protocols achieve limited security at the cost of performance degradation in terms of message overhead and latency time.

#### B. Cooperative Black Hole Attack

The blackhole attack has two phases. In the first phase, the malicious node exploits the ad hoc routing protocol such as AODV to advertise itself as having a valid route to a destination node, with the intention of intercepting packets, even though the route is spurious. In the second phase, the attacker node drops the intercepted packets without forwarding them. There is a more subtle form of this attack when an attacker node suppresses or modifies packets originating from some nodes, while leaving the data packets from other nodes unaffected. This makes it difficult for other nodes to detect the malicious node. In this work, however, a defense mechanism has been proposed against a cooperative blackhole attack in a MANET that relies on AODV routing protocol. Symbolic notations in Figure 1 are used in all the subsequent diagrams in the paper.

In the standard AODV protocol, when the source node *S* (Figure 2) wants to communicate with the destination node *D*, the source node *S* broadcasts the *RouteRequest* (RREQ) packet. Each neighboring active node updates its routing table with an entry for the source node *S*, and checks if it is the destination node or whether it has the current route to the destination node. If an intermediate node does not have the current route to the destination node, it updates the RREQ packet by increasing the hop count, and floods the network with the RREQ to the destination node *D* until it reaches node *D* or any other intermediate node that has the current route to *D*, as depicted in Figure 2.

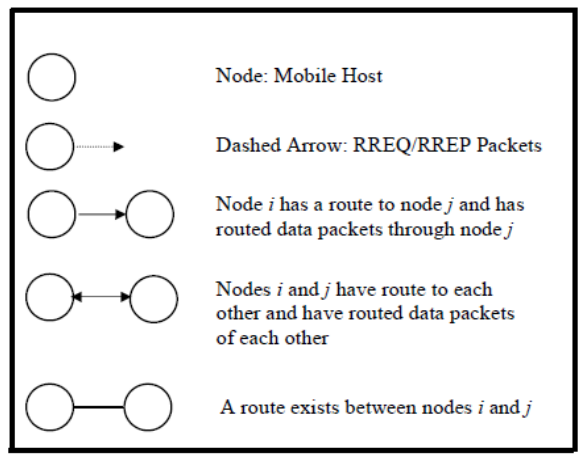

Figure 1. Symbolic notations used in diagrams

The destination node *D* or any intermediate node that has the current route to *D*, initiates a *RouteReply* (RREP) in the reverse direction, as depicted in Figure 3. Node *S* starts sending data packets to the neighboring node that responded first, and discards the other responses. This works fine when the network has no malicious nodes.

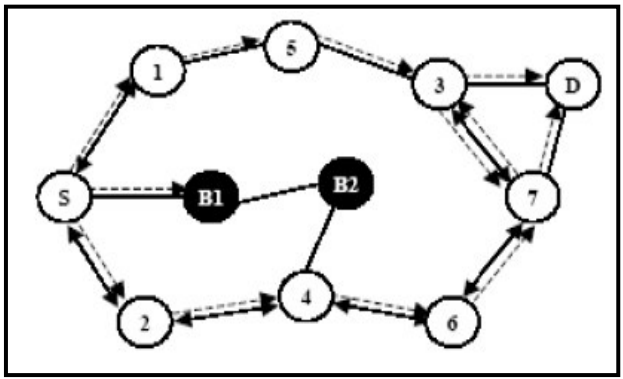

Figure 2. Network flooding by RREQ messages

In [2], authors have proposed a solution to identify and isolate a single blackhole node. However, the security threat arising out of the situation where multiple blackhole nodes act in coordination has not been addressed. For example, when multiple blackhole nodes are acting in coordination with each other, the first black hole node $B_1$ refers to one of its partners $B_2$ as the next hop, as depicted in Figure 2. In the mechanism propose in [2], the source node *S* sends a *FurtherRequest* (FRq) to $B_2$ through a different route ($S \rightarrow 2 \rightarrow 4 \rightarrow B_2$) other than via $B_1$. Node *S* asks $B_2$ if it has a route to node $B_1$ and a route to destination node *D*. Because $B_2$ is cooperating with $B_1$, its "*FurtherReply* (FRp)" will be "yes" to both the questions. According to the solution proposed in [2], node *S* starts sending the data packets assuming that the route $S$-$B_1$-$B_2$ is secure. However, in reality, the packets are intercepted and then dropped by node $B_1$ and the security of the network is compromised.

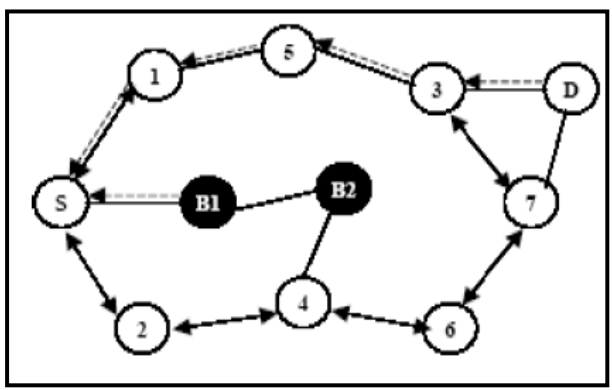

Figure 3. Propagation of RREP messages

## IV. THE PROPOSED ALGORITHM

Before you begin to format your paper, first write and save the content as a separate text file. In this, section the proposed mechanism for defending against a cooperative black hole attack is presented. The mechanism modifies the AODV protocol by introducing two concepts, (i) data routing information (DRI) table and (ii) cross checking.

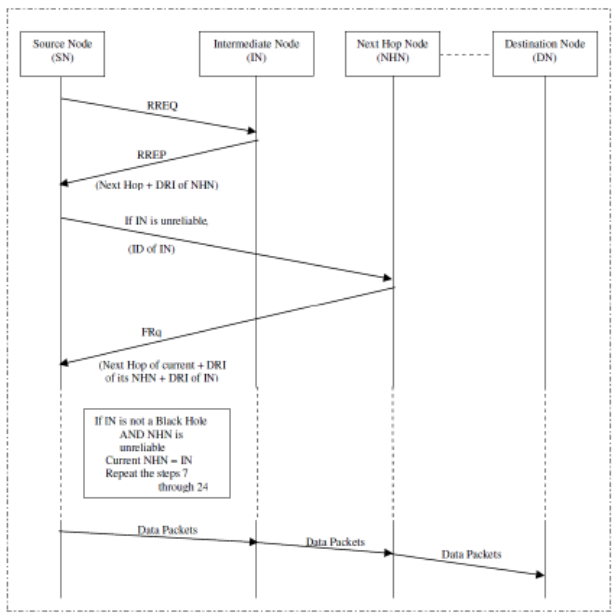

Figure 4. Modified AODV protocol to prevent cooperative black-hole attack

### A. Data Routing Information

In the proposed scheme, two bits of additional information are sent by the nodes that respond to the RREQ message of a source node during route discovery process. Each node maintains an additional data routing information (DRI) table. In the DRI table, the bit 1 stands for 'true' and the bit 0 stands for 'false'. The first bit 'From' stands for the information on routing data packet *from* the node (in the *Node* filed), while the second bit 'Through' stands for information on routing data packet through the node (in the *Node* field). With reference to the example depicted in Figure 3, a sample database maintained by node *4* is shown in Table 1. The entry 1 0 for node *3* implies that node *4* has routed data packets from *3*, but has not routed any data packets through *3* (before node *3* moved away from *4*). The entry 1 1 for node *6* implies that, node *4* has successfully routed data packets from and through node *6*. The entry 0 0 for node $B_2$ implies that, node *4* has not routed any data packets from or through $B_2$.

TABLE I. DRI TABLE MAINTAINED IN NODE 4

| Node # | Data Routing Information | |
|---|---|---|
| | *From* | *Through* |
| 3 | 1 | 0 |
| 6 | 1 | 1 |
| $B_2$ | 0 | 0 |
| 2 | 1 | 1 |

## B. Cross Checking

The proposed scheme relies on reliable nodes (nodes through which source has routed data previously and knows them to be trustworthy) to transfer data packets. The modified AODV protocol and the algorithm for the proposed mechanism are depicted in Figure 4 and Figure 5 respectively.

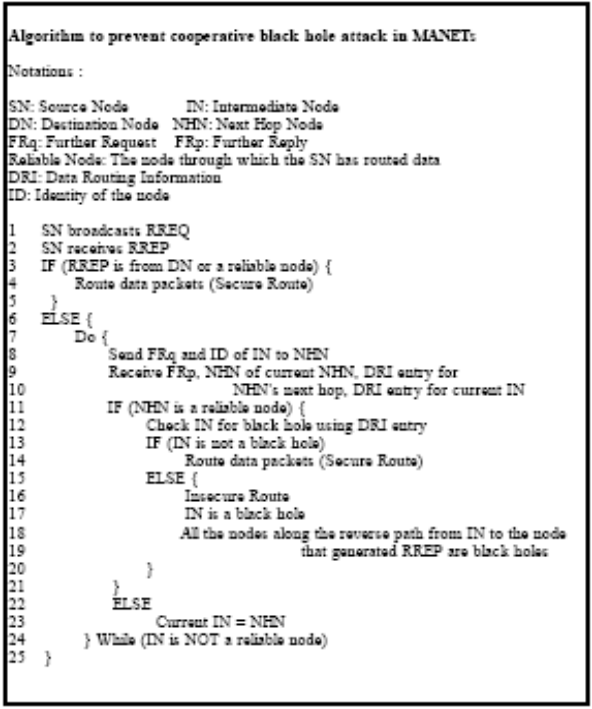

Figure 5. Modified AODV algorithm to detect co-operative black-hole attack

In the modified protocol, the *source node* (SN) broadcasts a RREQ message to discover a secure route to the destination node. The *intermediate node* (IN) that generates the RREP has to provide information regarding its *next-hop node* (NHN) and its DRI entry for that NHN.

Upon receiving the RREP message from IN, SN will check its own DRI table to see whether IN is its reliable node. If SN has used IN before for routing data packets, then IN is a reliable node for SN and SN starts routing data through IN. Otherwise, IN is unreliable and thus SN sends FRq message to NHN to check the identity of the IN, and asks NHN about the following information: (i) if IN has routed data packets through NHN, (ii) who is the current NHN's next hop to destination, and (iii) has the current NHN routed data through its own next hop. The NHN, in turn, responds with FRp message including the following responses: (i) DRI entry for IN, (ii) the information about its (NHN's) next hop node, and (iii) the DRI entry for its (NHN's) next hop. Based on the FRp message from NHN, SN checks whether NHN is reliable or not. If SN has routed data through NHN before, NHN is reliable; otherwise, NHN is unreliable for SN. If NHN is reliable, then SN will check whether IN is a blackhole or not. If the second bit of the DRI entry from the IN is equal to 1, i.e. IN has routed data *through* NHN, and the first bit of the DRI entry from the NHN is equal to 0 i.e. NHN has not routed data from IN, then IN is a blackhole. If IN is not a blackhole and NHN is a reliable node, then the route is secure, and SN will update its DRI entry for IN with 0 1, and starts routing data via IN. If IN is a blackhole, then SN identifies all the nodes along the reverse path from IN to the node that generated the RREP as blackhole nodes. Subsequently SN ignores any other RREP from the black-holes and broadcasts the list of cooperative black-holes in the network. If NHN is an unreliable node, SN treats current NHN as IN and sends FRq to the updated IN's next hop node and goes on in a loop from steps 7 through 24 in the algorithm depicted in Figure 5.

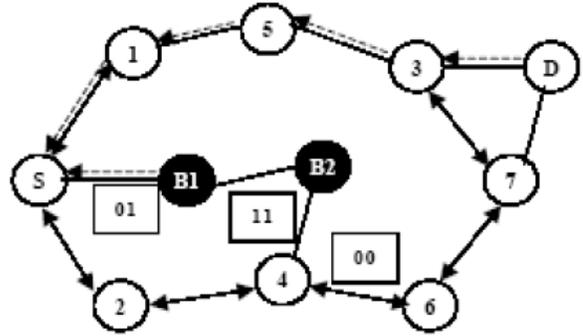

Figure 6. Detection of multiple black-hole nodes in one-time check

As an example, let's consider the network in Figure 6. When node $B_1$ responds to source node $S$ with RREP message, it provides its next hop node $B_2$ and DRI for the next hop (i.e. if $B_1$ has routed data packets through $B_2$). Here the blackhole node ($B_1$) lies about using the path by replying with the DRI value equal to 0 1. Upon receiving RREP message from $B_1$, the source node $S$ checks its own DRI table to see whether $B_1$ is a reliable node. Since $S$ has never sent any data through $B_1$ before, $B_1$ is not a reliable node to $S$. Therefore, $S$ sends FRq to $B_2$ via alternative path $S \rightarrow 2 \rightarrow 4 \rightarrow B_2$ and asks $B_2$ about three things: (i) whether $B_2$ has routed any data from $B_1$, (ii) who is $B_2$'s next hop, and (iii) whether $B_2$ has routed data packets through $B_2$'s next hop. Since $B_2$ is maliciously collaborating with $B_1$, it replies positively to all the three queries and gives node *6* (chosen randomly) as its next hop. When the source node contacts node *6* via alternative path $S \rightarrow 2 \rightarrow 4 \rightarrow 6$ to cross check the validity of the claims of node $B_2$, node *6* responds negatively. Since node *6* has neither a route to node $B_2$ nor it has received data packets from node $B_2$, the DRI value corresponding to $B_2$ as stored in node *6* is 0 0 as shown in Figure 6. Based on this information, node $S$ can infer that $B_2$ is a blackhole node. If node $B_1$ really had routed data packets through node $B_2$ before, it should have validated the node ($B_2$) before sending it. Now, since node $B_2$ is invalidated through node *6*, the source node $S$ infers that node $B_1$ is maliciously cooperating with node $B_2$. Hence both nodes $B_1$ and $B_2$ are marked as blackhole nodes and this information is propagated throughout leading to the revocation of their certificates. Subsequently *S* discards any further responses from $B_1$ or $B_2$ and looks from a valid alternative route to *D*.

The process of cross checking the intermediate nodes is a one-time procedure which should be affordable for the purpose of security. The cost of crosschecking the nodes can be minimized by allowing the nodes to share the DRI table of their trusted nodes with each other.

## V. SIMULATION AND RESULTS

The experiments for the evaluation of the proposed scheme have been carried out using the network simulator *ns-2*. The 802.11 MAC layer implemented in ns-2 is used for simulation. An improved version of random waypoint model is used as the model of node mobility [21]. Performances of the three protocols are evaluated: (i) Standard AODV protocol, (ii) AODV with two malicious nodes cooperating in a blackhole attack, and (iii) AODV with the proposed algorithm. The scenarios developed to carry out the tests use two parameters: (i) the mobility of the nodes and (ii) the number of active connections in the network. The chosen parameters for simulation are presented in Table II.

TABLE II. SIMULATION PARAMETERS

| Parameter | Value |
|---|---|
| Simulation duration | 1000 sec |
| Simulation area | 1000 * 1000 m |
| Number of mobile nodes | 30 |
| Transmission range | 200 m |
| Movement model | Random waypoint |
| Maximum speed | 5 – 20 m /sec |
| Traffic type | CBR (UDP) |
| Total number of flows | 15 |
| Packet rate | 2 packets / sec |
| Data payload | 512 bytes / packet |
| Number of malicious nodes | 2 |
| Host pause time | 10 sec |

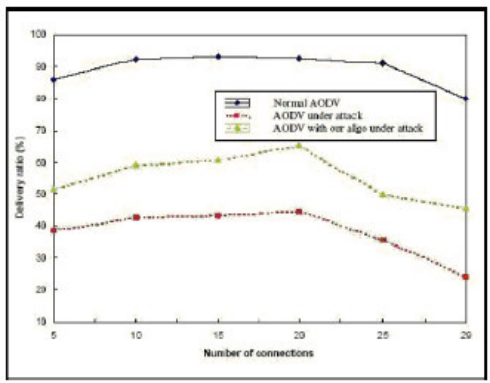

Figure 7. Packet delivery ratio vs. no. of active connections

Following metrics are chosen to evaluate the impact of the blackhole attack on the simulated network: (i) packet delivery ratio and (ii) false routing packets sent by the attacker nodes. These metrics were used to measure the severity of the attack and the improvement that the scheme manages to achieve during an active attack scenario. Every point in the graph (Figures 7-10) is an average of the values obtained after the experiment is repeated five times.

In Figure 7, the metric *packet delivery ratio* is plotted against the number of active connections. As the total number of mobile nodes in the network is 30, the maximum number of connections in the network is 29. A general observation from Figure 7 is that AODV achieves maximum delivery ratio with 10 to 25 active connections. However the delivery ratio falls slightly when the maximum number of connections (in this case 29) is established. The cooperative blackhole attack has a very severe impact as it decreases the delivery ratio to a value that is even less than half of the normal AODV delivery ratio. With the proposed scheme, the delivery ratio obtained is around 60%, which is a significant improvement over the degraded performance of the network under cooperative blackhole attack without any security mechanism.

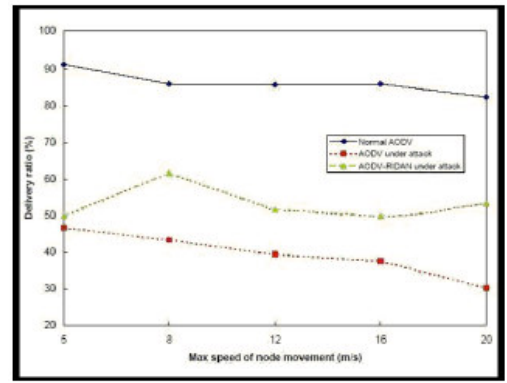

Figure 8. Delivery ratio vs. nodes' mobility

In Figure 8, packet delivery ratio is plotted against the mobility of the nodes. It is observed that AODV performs better for lower node mobility rates. The delivery rate starts dropping with increasing mobility of the nodes. The performance of the network significantly reduces when AODV is under the cooperative blackhole attack, and when the mobility of the nodes in the network increases. This behavior of the protocol is expected due to the following reason. With increasing mobility of the nodes the topology of the network changes faster, resulting in frequent route request generation. This gives an opportunity to a malicious node to send more false RREP packets. AODV under blackhole attack exhibits a decrease in delivery ratio to 38%. The proposed algorithm increases the delivery ratio to 55%, resulting in an average improvement of 17%.

The second metric that is used in the evaluation of the attack is the number of false packets sent by the attacker nodes versus the number of active connections in the network. This metric has been used to examine the overhead of the blackhole attack. From Figure 9, it is observed that the average number of false RREPs sent by the malicious nodes in all the experiments conducted was 2056, and the number of nodes that inserted the false route into their routing table was 22 out of the total 30 nodes. Figure 10 shows that with the increase in the mobility, the number of false RREP packets sent by the malicious nodes also increases.

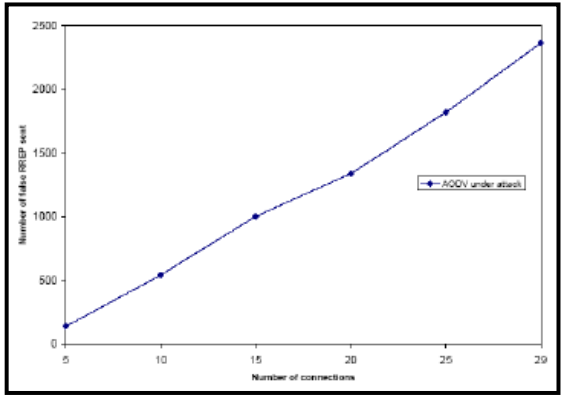

Figure 9. No. of false replies from attacker nodes vs. no. of connections

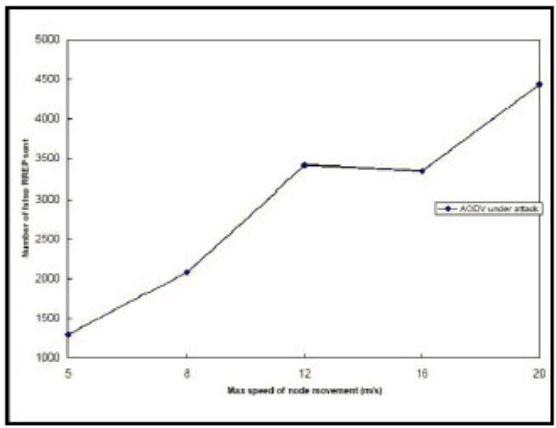

Figure 10. No. of false replies sent by attacker nodes vs. nodes' mobility

In order to further study the performance of the scheme under varying number of black-hole nodes, the percentages of black-hole nodes are varied from 0% to 20% and performance of the protocol is measured with respect to four parameters: (i) false positive rate, i.e., the probability of incorrectly identifying an honest node as malicious, (ii) false negative rate, i.e., the probability of failure in detecting a malicious node, (iii) data packet delivery ratio, i.e., the percentage of data packets that is successfully delivered, and (iv) communication overhead due to control packets of the security algorithm.

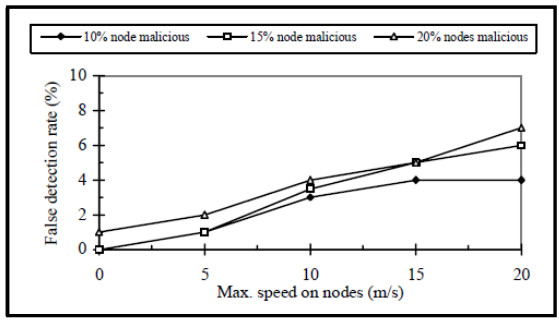

Figure 11. False positive rate vs. nodes' mobility

Figure 11 shows how false positive rate varies with the mobility of the nodes for different percentages of black-hole nodes in the network. The maximum value of observed false positive rate is found to be 7%. The false positive rate increases as the nodes move faster. If a node constantly moves at a high speed, it can gather only partial information about its transmissions with the current neighbors. As a result, it is more likely to make errors in detection.

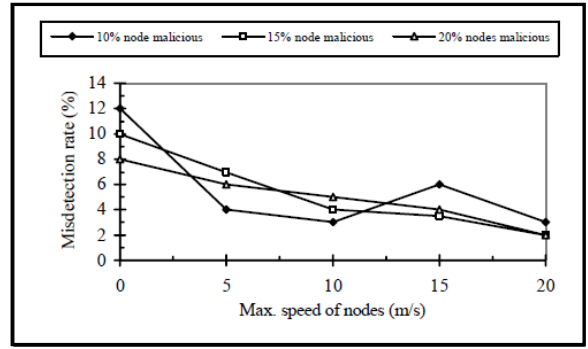

Figure 12. Misdetection (false negative) rate vs. nodes' mobility

Figure 12 depicts the variation of false negative rate with nodes' mobility for different percentages of black-hole nodes. It is seen that false negative rate is maximum (12%) in a static network which starts dropping with the mobility of the nodes. This is because in a static network if a black hole remains in a sparsely populated region, its neighbors may not be able to punish it since there may not be requisite k number of nodes to arrive at the consensus. On the contrary, in a mobile network, the mobility increase the probability that other nodes roam into the region of black hole node(s), or the black hole node(s) enters into a densely populated region. As a result, it is less likely that a black-hole would be able to escape without being detected.

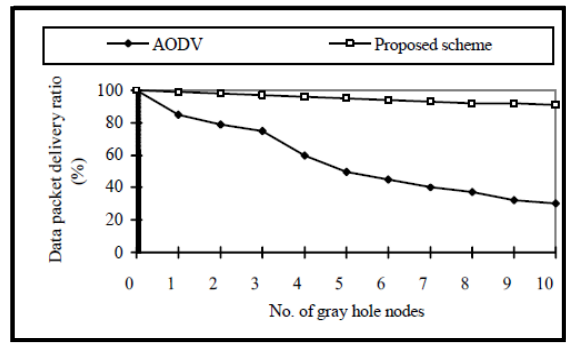

Figure 13. Data packet delivery rate vs. no. of black-holes

Figure 13 shows how the data packet delivery ratio varies with respect to the number of black-hole nodes for the base AODV protocol, and for the proposed secure routing protocol. It is observed that even when 20% of the nodes in the network are malicious black-holes, the percentage of packets successfully delivered is more than 90%, if the proposed security protocol is utilized. However, 100% packet delivery ratio is not achieved even with the proposed security protocol. A careful analysis of the trace files has shown that most of the packet loss occurs during the detection and reaction phase of the algorithm.

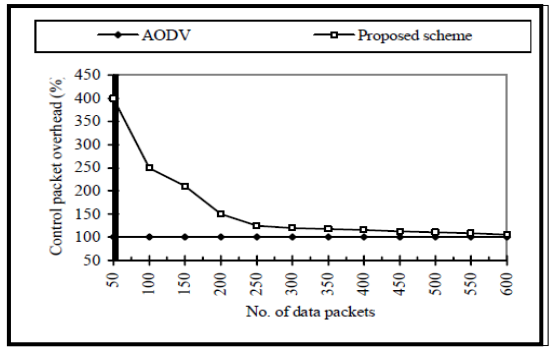

Figure 14. Control packet overhead due to different no. of data packets

Figure 14 shows the communication overhead due the control packet exchanges in the security protocol. It has been observed that with the increase in number of black-hole nodes, the overhead increases. The result for the worst case has been reported in Figure 14, in which case the percentage of malicious node is 20%. The communication overhead is as shown as the percentage of the number of control packets required to route different number of data packets in the network. The performance of the plain AODV is taken as the baseline. It is observed that the overhead due to the proposed security mechanism decreases as the number of data packets transmitted is increased. This clearly demonstrates that the scheme is efficient and has a low communication overhead.

VI. CONCLUSION

In this paper, routing security issues in MANETs are discussed in general, and in particular the cooperative blackhole attack has been described in detail. A security protocol has been proposed that can be utilized to identify multiple blackhole nodes in a MANET and thereby identify a secure routing path from a source node to a destination node avoiding the blackhole nodes. The proposed scheme has been evaluated by implementing it in the network simulator *ns-2*, and the results demonstrate the effectiveness of the mechanism. As a future scope of work, the proposed security mechanism may be extended so that it can defend against other attacks like resource consumption attack and packet dropping attack, grey-hole attack etc.